\documentclass[pre,twocolumn,aps,superscriptaddress,showpacs,floatfix]{revtex4}
\usepackage{graphicx}
\usepackage{dcolumn}
\usepackage{bm}
\usepackage{amsmath}
\begin{document}
\title{ Collective shuttling of attracting particles in asymmetric narrow channels}
\author{A. Pototsky}
\affiliation{Department of Physics, Loughborough University,
Loughborough LE11 3TU, United Kingdom}
\author{A.J. Archer}
\affiliation{Department of Mathematical Sciences, Loughborough University,
Loughborough LE11 3TU, United Kingdom}
\author{M. Bestehorn}
\affiliation{Lehrstuhl f{\"u}r Theoretische Physik II,
 Brandenburgische Technische Universit\"at Cottbus,
 Erich-Weinert-Stra{\ss}e 1, D-03046 Cottbus, Germany}

\author{D. Merkt}
\affiliation{Lehrstuhl f{\"u}r Theoretische Physik II,
 Brandenburgische Technische Universit\"at Cottbus,
 Erich-Weinert-Stra{\ss}e 1, D-03046 Cottbus, Germany}
\author{S. Savel'ev}
\affiliation{Department of Physics, Loughborough University,
Loughborough LE11 3TU, United Kingdom}
\author{F. Marchesoni}
\affiliation{Dipartimento di Fisica, Universit\`a di Camerino,
I-62032 Camerino, Italy}

\begin{abstract} The rectification of a single file of attracting particles subjected to a low frequency ac drive is proposed as a working mechanism for particle shuttling in an asymmetric narrow channel. Increasing the particle attraction results in the file condensing, as signalled by the dramatic enhancement of the net particle current. Magnitude and direction of the current become extremely sensitive to the actual size of the condensate, which can then be made to shuttle between two docking stations, transporting particles in one direction, with an efficiency much larger than conventional diffusive models predict.

\end{abstract}

 \maketitle

Diffusion and directed transport of finite size interacting Brownian particles in constrained geometries are issues of growing relevance in soft matter physics \cite{RMP09}. In particular, the dynamics of one-dimensional (1d) gases, or single files, has found wide applications, ranging from the modeling of natural processes, such as in molecular motors and ion channels \cite{hille01}, to the design of artificial devices, involving, e.g., engineered porous materials \cite{kaerger92}, colloids \cite{wei00,lutz04}, and superconducting vortices \cite{wambaugh}.

A key feature, which is intrinsic to 1d files in both natural and artificial channels, is attractive particle interactions, whose origin has been variously tracked down to excluded volume effects \cite{vicsek95}, Van der Waals forces \cite{coupier07}, or the presence of additional passive molecules, like in colloid-polymer mixtures \cite{strander}. It has been recognized that pair attractive forces can induce the formation of large molecular clusters and that a  dramatic increase in the overall file diffusion and mobility can ensue as a non-commensuration effect of condensate size and channel spatial periodicity \cite{sholl}. 
More generally, the incommensurability between the lateral dimensions of a molecule and a catalyst is responsible for the ``shape selectivity" of many a biological channel \cite{smit}.

In this Letter we combine three different and important mechanisms, namely, (i) condensation in single files, (ii) 1d transport and diffusion, and (iii) the ratchet effect and its applications, to outline the blueprint for a new class of nanoparticle shuttles. We show that Brownian particles moving along a narrow channel with a ratchet-like profile, subjected to a low frequency center-symmetric ac drive, can condense to a cluster when the temperature is lower than a certain threshold. This results in the net time-averaged particle current $\langle J \rangle$ being much larger (up to over one order of magnitude) than in the presence of weak or no attraction, when the particles are evenly distributed throughout the channel. Moreover, $\langle J \rangle$ strongly depends on the ratio $hN/L$, where $h$ is the particle length, $N$ is the number of particles in the system, and $L$ is the ratchet spatial period. Thus, in contrast to earlier reports, the ratchet current through the channel not only is strongly enhanced by condensation, but it can also be inverted by adding or subtracting particles to the cluster (for appropriate $h/L$, just one!). This suggests a new concept of collective particle shuttle, where the motion of the cluster gets reversed any time, say, one particle is unloaded (uploaded) at a docking station (respectively, a sink or a source). Besides applying to the diffusion of long molecular chains in zeolites and the shape selective control of catalytic reactions in living cells, collective shuttling can also answer the longstanding question as why proton and electron biological pumps work much more effectively than predicted by purely diffusive shuttle models \cite{smirnov}.

\begin{figure}[btp]
\centering
\includegraphics[width=0.48\textwidth]{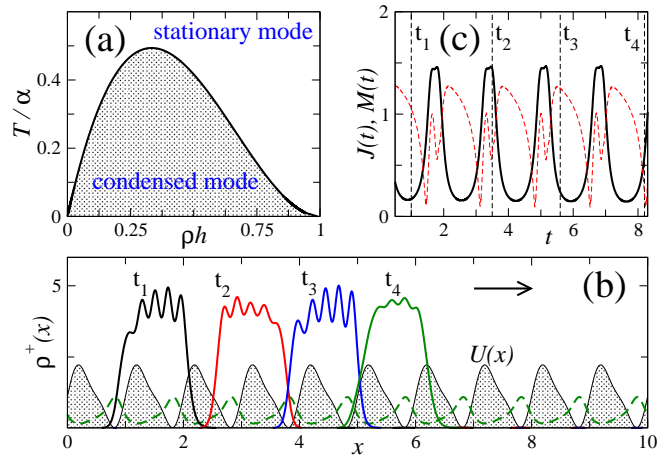}
\caption{(Color online) (a) Phase diagram of a infinitely long single-file with $\lambda=3$, $T=1$, $h=0.2$. In the shaded region configurations with uniform density are linearly unstable due to particle attraction. (b) Density profiles $\rho^{+}(x,t)$ in a finite system of length $S=10$ and with $N=5$ particles, obtained by solving Eq.\ \eqref{FP}. The dashed line represents the diffusion dominated case at $\alpha=0$. The density profiles labeled $t_1$--$t_4$ are four snapshots of the traveling condensed mode (the arrow indicates the direction of motion) for $A=1$, $\alpha=10$ and the remaining parameters as in (a). $U(x)$ is also displayed for convenience. (c) Instantaneous current $J(t)$ (solid line) and form factor $M(t)$ (dashed line) vs.\ time for the simulation parameters of panel (b). The vertical dashed lines denote the snapshot times $t_1$--$t_4$ also used in (b).\label{F1}}
\end{figure}

To demonstrate how ratchet effect can be controlled in the presence of attracting particles, we study a model system consisting of $N$ Brownian particles moving along a circular 1d channel. The particles interact via the pair potential $w(x_{ij})=w_{\rm hr}(x_{ij}) + w_{\rm at}(x_{ij})$, which consists of a hard-core repulsive, $w_{\rm hr}$, and a long range attractive part, $w_{\rm at}$, with $x_{ij}=|x_i-x_j|$ denoting the distance between the centers of particles $i$ and $j$. In the following we assume $w_{\rm at}(x_{ij})=-\alpha\exp(-\lambda x_{ij})$, where  $\lambda^{-1}$ and $\alpha$ characterize, respectively, the range and the strength of the pair attraction. The dynamics of the particles is governed by a set of coupled overdamped stochastic Langevin equations (in rescaled units)
\begin{eqnarray}
\label{langevin}
\frac{d x_i}{d t} = -\frac{\partial \Phi(\{x_j\},t)}{\partial x_i} +\sqrt{2T} \xi_i(t).
\end{eqnarray}
Here $\xi_i(t)$ is a stochastic white noise with zero mean and autocorrelation function $\langle \xi_i(t) \xi_i(t')\rangle=\delta(t-t')$, $T$ is the temperature, and $\Phi(\{x_j\},t)$ is the total potential energy of the system, $\Phi(\{x_j\},t)=\sum_i[U(x_i)-F(t)x_i]+\frac{1}{2}\sum_{j \neq i}\sum_i w(| x_i -x_j |)$, where $U(x)$ represents the ratchet potential and $F(t)$ an external ac drive. We choose a biharmonic function with period $L=1$ for the ratchet potential, $U(x)=\sin{(2\pi x)}+0.25 \sin{(4\pi x)}$ [see Fig. \ref{F1}(b)], and a square-wave with amplitude $A$ for $F(t)$. Finally, we set periodic boundary conditions at the endpoints of the channel, the length of which is a multiple of $L$, $S=ML$. In doing so, we implicitly assume that $hN \leq ML$ and that $w_{\rm at}(x)$ effectively vanishes for distances comparable with $S$, $w(S) \approx 0$.

To investigate the non-equilibrium properties of the system we numerically integrate the Langevin equations (\ref{langevin}). 
For computational reasons, in our simulations we replace the hard-core potential $w_{\rm hr}$ by the soft-core power-law potential $w_{\rm s}(x_{ij}) = \epsilon (h_*/x_{ij})^{19}$. For fixed $\epsilon$ and $h_*$ the effective hard core length $h_{\rm eff}$ of these particles is a function of $\alpha$ and $T$  \cite{barker76}; in our simulations $\epsilon=0.01$ and $h_*=0.2$ correspond to $h_{\rm eff}\simeq 0.16$ for $\alpha=10$.
We also study the system in the framework of the dynamical density functional theory (DDFT) \cite{marini99}. DDFT approximates the Fokker-Planck equation for the one-particle density distribution $\rho(x,t)$ to:
\begin{eqnarray}
\label{FP}
\frac{\partial \rho(x,t)}{\partial t} = \frac{\partial}{\partial x}\left[ \rho(x,t)\frac{\partial }{\partial x}\frac{\delta F[\rho(x,t)]}{\delta \rho(x,t)}\right],
\end{eqnarray}
where $F[\rho]$ is the Helmholtz free energy functional obtained from the equilibrium density functional theory \cite{evans1992fif, hansen2006tsl}. For hard rods with no attraction, $w_{\rm at}=0$, an exact analytical derivation yields the equilibrium free energy term \cite{percus76} $F_{\rm hr} = T\int_{-\infty}^{\infty} dx\,\rho(x,t)(\ln \rho(x,t)-1)+\frac{1}{2}\int_{-\infty}^{\infty} dx\,\phi[\rho(x,t)][\rho(x+h/2,t)+\rho(x-h/2,t)]$, with $\phi[\rho(x,t)] = -T\ln{[1-\eta(x,t)]}$ and $\eta(x,t) = \int_{x-h/2}^{x+h/2}dy\,\rho(y,t)$. Next, by adding a mean-field  term to account for the particle attraction \cite{evans1992fif}, we obtain $F[\rho] = F_{\rm hr} + F_{\rm at}$, where $F_{\rm at} = \frac{1}{2}\int dx \int dy\,w_{\rm at}(\mid x-y \mid)\rho(x)\rho(y)$.
Moreover, in the present case the function $\rho(x,t)$ in Eq.\,(\ref{FP}) obeys periodic boundary conditions in space with period equal to the system size, namely, upon setting the center of the system at $x=0$, $\rho(-S/2,t) = \rho(S/2,t)$.
Finally, on rewriting Eq.\,(\ref{FP}) as $\partial \rho(x,t) / \partial t = -\partial J(x,t)/ \partial x$, we introduce the instantaneous particle current density, $J(x,t)$. In the presence of the dc drives $F=\pm A$, the unidirectional currents $J^{\pm}$ are computed as the time average of the instantaneous current divided by the number of particles $N$, i.e.\ $J^{\pm} = \lim_{t\rightarrow \infty}(1/Nt)\int_{t^\prime}^{t^\prime+t}\,d\tau\,\int_{-S/2}^{S/2}J(x,\tau)\,dx$.
To extract $J^{\pm}$ from Eq.\,(\ref{FP}), we perform a Fourier mode expansion of $\rho(x,t)$ and then apply a standard continuation technique \cite{AUTO}.

{\it Two transport modes.} When the pair attraction is sufficiently strong, the particles in the channel condense to form a compact cluster, which behaves like a single composite object. This phenomenon may be understood by considering a free single-file with $F(t)=0$ and $U(x)=0$. The equilibrium properties of such a homogeneous system are well known \cite{hansen2006tsl,standard_ref} in the thermodynamic limit ($N,S \to \infty$ with constant density $\rho=N/S$). As the attraction strength, $\alpha$, is increased, the particles tend to group together. Above some threshold value of $\alpha$, in 2d and 3d systems the homogeneous state undergoes a phase transition to a state characterized by the coexistence of a low-density gas and a high-density liquid phase. In 1d systems, particles also tend to cluster together on raising $\alpha$, but no true phase transition can be defined \cite{standard_ref}.

Within our mean-field DDFT treatment, the homogeneous one-body density $\rho(x)$ becomes linearly unstable when the free energy per volume $F[\rho]/S$ turns concave, i.e. for $\delta^2 F/\delta \rho^2 <0$. As a consequence, inside the shaded (spinodal) region of Fig.\,\ref{F1}(a), the attractive interactions are strong enough to overcome the stabilizing action of diffusion, thus making an initially homogeneous single-file unstable. For a system with finite size $S$ and finite number of particles $N$, such an instability signals the transition from an homogeneous state with evenly distributed particles, to a condensed state characterized by a single cluster of approximate length $hN$ and diffusing with effective temperature $T/N$.

In the presence of periodic pinning $U(x)$ and non-zero dc drive, condensation explains the emergence of two different transport modes: a stationary homogeneous mode and a condensed traveling mode. To illustrate the transition between these two modes, we set $T=1$, $A=1$ and gradually increase $\alpha$. For small $\alpha$, the density profiles $\rho^{\pm}(x)$, corresponding to $F=\pm 1$, are pinned by $U(x)$, namely, are stationary functions with spatial period $L=1$. This means that the particles are evenly distributed among the local minima of $U(x)$, jumping occasionally from one well to the next in the direction of the drive, as illustrated in Fig.\,\ref{F1}(b) for $F=1$. This is the standard ratchet mechanism \cite{RMP09}, which dominates at high temperatures or, equivalently, for weak interactions or dilute single-files. On increasing $\alpha$ this mode becomes unstable and is eventually replaced by a condensed traveling mode, resembling a wave packet of width $hN$ traveling in the direction of the drive. Different phases of the traveling mode are shown in Fig.\,\ref{F1}(b) at four different time instants, $t_1<t_2<t_3<t_4$. As the $N=5$ particle cluster moves over the barrier of the pinning potential $U(x)$, the interparticle distance gets modulated in time, giving rise to spatio-temporal cluster oscillations. The period of these oscillations coincides with the time needed for the cluster to move across a unit cell of $U(x)$. This barrier crossing mechanism is better visualized in Fig.\,\ref{F1}(c), where the instantaneous velocity of a cluster particle, $J(t)=(1/N)\int_{-S/2}^{S/2}J(x,t)\,dx$, is contrasted with the form factor, $M(t)=\rho^{+}_{\rm max} - \rho^{+}_{\rm min}$, introduced to quantify the spatial oscillations of the cluster. Here $\rho^{+}_{\rm max}$ is the absolute maximum and $\rho^{+}_{\rm min}$ the largest of the local minima of the instantaneous cluster density $\rho^{+}(x,t)$.

\begin{figure}[btp]
\centering
\includegraphics[width=1.\columnwidth]{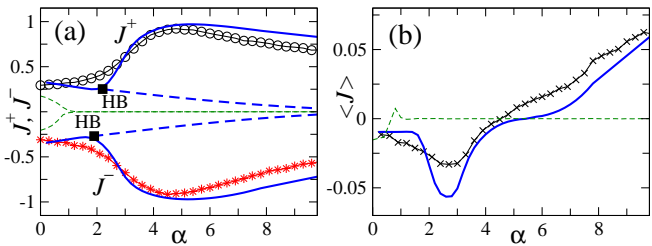}
\caption{(Color online)  (a) Unidirectional currents $J^{\pm}$ vs. $\alpha$, for $T=0.5$, $A=1$, $N=5$, $M=5$, $\lambda=3$, and $h=0.16$. The solid and dashed lines are, respectively, the stable and unstable solutions of Eq.\,(\ref{FP}); symbols represent the corresponding simulation data. The Hopf bifurcations are marked by a square labeled HB. $J^{\pm}$ for $h=0$ are displayed for comparison (dotted lines). (b) Zero-frequency rectification currents, $\langle J \rangle$, obtained from the data in (a). The positive $\alpha \to \infty$ asymptotes coincide with the point $hN=0.85$ in Fig. \ref{F3}(c). \label{F2}}
\end{figure}

{\it Current enhancement.} The efficiency of particle transport along the channel is characterized by the unidirectional currents $J^{\pm}$ driven by $F=\pm A$. Correspondingly, the rectification current, $\langle J \rangle$, induced by a square-wave ac drive $F(t)$ of amplitude $A$ and vanishingly low frequency, is simply the average $\langle J \rangle=\frac{1}{2}(J^{+}+J^{-})$. In Fig.\,\ref{F2}(a) we compare our DDFT results for the two currents $J^{\pm}$ as a function of  $\alpha$ (solid lines), with the simulation outcome (symbols).
As expected, for small $\alpha$ the long time limit solutions of Eq.\,(\ref{FP}) are stationary functions with spatial period $L$. The branch of the $L$-periodic stationary solutions becomes unstable via a Hopf bifurcation (HB) at $\alpha_{HB} \simeq2$; see Fig.\,\ref{F2}(a). For $\alpha$ above the HB threshold $\alpha_{HB}$, a new branch of stable time-periodic solutions appears. The transition from the stationary to the time-periodic regime corresponds to the transition from the stationary to the traveling condensed mode. At the onset of the traveling mode, immediately to the right of the HB point, both $|J^{\pm}|$ increase sharply with $\alpha$, well above their zero attraction values, until, for $\alpha \approx 5$, they approach the expected upper bound, $|J^{\pm}|=A$ (unpinned single-file regime \cite{taloni}).

The transport enhancement in the traveling mode is not solely a consequence of the particle attraction: by preventing condensation to a vanishingly small volume, hard-core particle repulsion also plays a crucial role. If one sets $h=0$, no enhancement of the particle mobility with increasing $\alpha$ occurs
[dotted lines in Fig.\,\ref{F2}(a)]; on the contrary, for pointlike particles $J^{\pm}$ drop by as much as one order of magnitude after condensation sets in, as already pointed out in Ref.\ \cite{sav04}.

The difference between the unidirectional currents $J^{\pm}$ depends both on the interaction strength $\alpha$ and on the particle length $h$. Indeed, the $\alpha$ dependence of the zero-frequency ratchet current, $\langle J \rangle=\frac{1}{2}(J^{+}+J^{-})$, may widely vary with $h$. In the example of Fig.\,\ref{F2}(b), $|\langle J \rangle|$ increases rapidly, retaining its initial ($\alpha=0$) negative sign. However, for $\alpha>5$ it reverses sign and eventually levels off with even larger a modulus. Note that the existence of a zero-crossing in the curve $\langle J \rangle$ versus $\alpha$ is determined by the sign of the curve asymptote for $\alpha \to \infty$. In this limit $\langle J \rangle$ oscillates with $h$ as one better sees by investigating our model in the strong attraction approximation.

\begin{figure}[btp]
\centering
\includegraphics[width=1.\columnwidth]{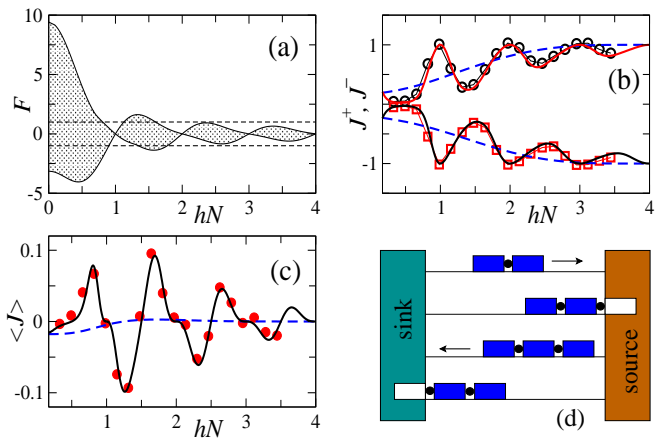}
\caption{(Color online) (a) Pinning $(F,hN)$ region (shaded) for the condensed mode in the limit $\alpha\to \infty$, Eq.\,(\ref{langevin2}), and $T=0$. The horizontal lines correspond to $F=\pm1$ used in the simulations. (b) Solid lines: $J^{\pm}$ vs.\ $hN$, obtained integrating Eq.\,(\ref{langevin2}) for $h=0.16$; symbols: data points obtained via simulation of Eqs.\,(\ref{langevin}) with $\alpha=10$, $\lambda=3$, $h_{\rm eff}\simeq0.16$ and varying $N$; dashed lines: $J^{\pm}$ vs.\ $hN$ computed from Eq.\,(\ref{FP}) with $\alpha=0$. Other parameters are $A=1$, $T=0.5$, and $M=10$. (c) The zero-frequency rectification currents $\langle J \rangle$ vs.\ $hN$ computed using the data in (b). A qualitatively similar plot is obtained on varying $h$ for fixed $N$. The current peaks should be compared with $J_{max}=0.5$. (d) Sketch of the molecular shuttle described in the text. \label{F3}}
\end{figure}

{\it Cluster size and current reversals.} In the limit $\alpha \to \infty$, when particle attraction dominates over thermal diffusion and static pinning, Eqs.\,(\ref{langevin}) can be reduced to a single equation describing the evolution of the center of mass, $y=(1/N)\sum_i x_i$, of the particle condensate,
\begin{eqnarray}
\label{langevin2}
\frac{d y}{d t} = -\frac{U(y+hN)-U(y)}{hN} +F(t)+ \sqrt{\frac{2T}{N}}\xi(t),
\end{eqnarray}
where $\xi(t)$ has the same statistics as $\xi_i(t)$ in Eqs.\,(\ref{langevin}). This equation helps clarify the role of the particle size, $h$, and the number, N, of particle in the cluster in the depinning of the condensed mode. By inspecting Eq.\,(\ref{langevin2}), we notice that for large $N$ diffusion becomes negligible, and the unidirectional currents $J^{\pm}$ set in only when the drive amplitude $A$ overcomes the pinning force exerted by $U(x)$, either from the right, $F_R$, or from the left, $F_L$ (both $F_{R,L}$ are functions of $hN$). Note that for a single particle, $N=1$, or for $hN \to 0$, $F_R=3\pi$ and $F_L=3\pi/2$ \cite{RMP09}. Upon neglecting thermal fluctuations, $J^{\pm}$ and the zero-frequency rectification current, $\langle J \rangle$, can both be computed analytically from Eq.\,(\ref{langevin2}) 
\cite {RMP09} as functions of $hN$. For $hN$ and $F$ chosen within the shaded region of Fig.\,\ref{F3}(a) the condensed mode is pinned. Nonzero currents thus emerge only due to thermal fluctuations with effective temperature $T/N$.  Most remarkably, for appropriate combinations of $h$ and $N$, one can achieve complete locking of the condensed mode in one direction, but not in the other \cite{sav04}, leading to the observation that for a certain value of $A$ the modulus of $\langle J \rangle$ cannot exceed $J_{max}=A/2$. This is a benchmark against which to compare the efficiency of single-file rectification.

For finite $T$, $J^{\pm}$ depend on $h$ and $N$ separately. In Fig.\,\ref{F3}(b) we plot $J^{\pm}$ versus $N$ in three different regimes: in the $\alpha \to \infty$ limit described by Eq.\,(\ref{langevin2}) with $h=0.16$ (solid lines); from numerical simulation of the original system, Eq.\,(\ref{langevin}) with strong pair attraction, $\alpha=10$, $\lambda=3$, $M=10$ and $h_{\rm eff}\simeq 0.16$ (symbols); and from the DDFT approximation, Eq.\,(\ref{FP}), for $\alpha=0$, $h=0.16$ and $M=4$ (dashed lines). All other parameters, $T=0.5$, $A=1$, are the same. The corresponding zero-frequency rectification currents, $\langle J \rangle$, are displayed in Fig.\,\ref{F3}(c). Similar behaviors can be obtained by varying $h$  at constant $N$. In conclusion, on increasing the size of the condensed cluster our model exhibits: (i) an oscillatory growth of $J^{\pm}$, which eventually tend to the respective (unpinned particle) limits $J^{\pm}=\pm 1$, and (ii) sign reversals of $\langle J \rangle$ for $hN$ an integer multiple of $L/2$.
The qualitative explanation of this behavior is simple for $\alpha \to \infty$, when the single-file of particles can be regarded as a continuous rod of length $hN$. The total pinning force exerted by $U(x)$ on such a traveling rod vanishes for $hN=nL$, with $n$ a positive integer (zero effective pinning), whereas its intensity is maximum for $hN \approx (2n-1)L/2$. Correspondingly, $J^{\pm}=\pm 1$ for $hN$ an integer multiple of $L$ [Fig.\,\ref{F3}(b)] and $\langle J \rangle=0$ for $hN$ an integer multiple of $L/2$ [Fig.\,\ref{F3}(c)]. For finite $\alpha$ the effective rod is somewhat larger than $hN$, so that zeros of $\langle J \rangle$ may occur for $hN <1$, like in Fig.\,\ref{F2}(b).

{\it Collective shuttles.} Particle size and number can be used to selectively control the sign and magnitude of the net current. This concept is illustrated and applied in the toy molecular shuttle, sketched in Fig.\ \ref{F3}(d), where a cluster of $N=2$ rod-shaped molecular units, each of length $h=0.4$ (in dimensionless units), moves to the right [Fig.\,\ref{F3}(c)] until it docks at the loading station at the exit of the channel (source). After binding to an additional molecular unit, the enlarged cluster now reverses its drift velocity [Fig.\,\ref{F3}(c)], thus dragging its cargo to the unloading station at the left exit of the channel (sink), where it releases the leftmost unit, so that the process can repeat itself. A steady molecular flux is thus established along the channel between the source and the sink, more efficiently than by mere single molecule (or, worse, single-file) diffusion. Consequently, our shuttle concept provides a much more efficient mechanism for ion  transport in bio pumps than standard diffusion shuttles \cite{smirnov}. Finally, we stress that high-performance collective shuttles only work with entrained particles of finite-size; for pointlike particles the resulting condensate cluster would have zero width and the file current would be suppressed.


AJA gratefully acknowledges support from RCUK. This work was partly supported by the HPC-Europa2 Transnational Access Programme.


\begin{thebibliography}{23}
\expandafter\ifx\csname natexlab\endcsname\relax\def\natexlab#1{#1}\fi
\expandafter\ifx\csname bibnamefont\endcsname\relax
  \def\bibnamefont#1{#1}\fi
\expandafter\ifx\csname bibfnamefont\endcsname\relax
  \def\bibfnamefont#1{#1}\fi
\expandafter\ifx\csname citenamefont\endcsname\relax
  \def\citenamefont#1{#1}\fi
\expandafter\ifx\csname url\endcsname\relax
  \def\url#1{\texttt{#1}}\fi
\expandafter\ifx\csname urlprefix\endcsname\relax\def\urlprefix{URL }\fi
\providecommand{\bibinfo}[2]{#2}
\providecommand{\eprint}[2][]{\url{#2}}

\bibitem{RMP09} F. Marchesoni and P. H\"anggi, Rev. Mod. Phys. {\bf 81}, 387 (2009).
\bibitem[{\citenamefont{Hille}(2001)}]{hille01}
\bibinfo{author}{\bibfnamefont{B.}~\bibnamefont{Hille}},
  \emph{\bibinfo{title}{Channels of Excitable Membranes}}
  (\bibinfo{publisher}{Sinauer Asc., Sunderland}, \bibinfo{year}{2001}).

\bibitem[{\citenamefont{K{\"a}rger and Ruthven}(1992)}]{kaerger92}
\bibinfo{author}{\bibfnamefont{J.}~\bibnamefont{K{\"a}rger}} \bibnamefont{and}
  \bibinfo{author}{\bibfnamefont{D.~M.} \bibnamefont{Ruthven}},
  \emph{\bibinfo{title}{Diffusion in Zeolites and Other Microporous Solids}}
  (\bibinfo{publisher}{Wiley, New York}, \bibinfo{year}{1992}).

\bibitem{wei00} Q.H. Wei {\it et al.}, Science {\bf 287}, 625 (2000); B. Cui, H. Diamant and B. Lin, Phys. Rev. Lett. {\bf 89}, 188302 (2002).

\bibitem[{\citenamefont{Lutz et~al.}(2004)\citenamefont{Lutz, Kollmann, and
  Bechinger}}]{lutz04}
\bibinfo{author}{\bibfnamefont{C.}~\bibnamefont{Lutz}~{\it et~al.}},
  \bibinfo{journal}{Phys. Rev. Lett.} \textbf{\bibinfo{volume}{93}},
  \bibinfo{pages}{026001} (\bibinfo{year}{2004}).

\bibitem{wambaugh} J. F. Wambaugh {\it et al.}. Phys. Rev. Lett. {\bf 83}, 5106 (1999).

\bibitem[{\citenamefont{Derenyi and Vicsek}(1995)}]{vicsek95}
\bibinfo{author}{\bibfnamefont{I.}~\bibnamefont{Derenyi}} \bibnamefont{and}
  \bibinfo{author}{\bibfnamefont{T.}~\bibnamefont{Vicsek}},
  \bibinfo{journal}{Phys. Rev. Lett.} \textbf{\bibinfo{volume}{75}},
  \bibinfo{pages}{374} (\bibinfo{year}{1995}).

\bibitem[{\citenamefont{Coupier et~al.}(2007)\citenamefont{Coupier, Sain~Jean, and Guthmann}}]{coupier07}
\bibinfo{author}{\bibfnamefont{G.}~\bibnamefont{Coupier}~{\it et~al.}},
  \bibinfo{journal}{Euro. Phys. Lett.} \textbf{\bibinfo{volume}{77}},
  \bibinfo{pages}{60001} (\bibinfo{year}{2007}).

\bibitem{strander} A. Stradner {\it et al.}, Nature {\bf 432}, 492 (2004).

\bibitem{sholl} D.S. Sholl and K.A. Fichthorn, Phys. Rev. Lett. {\bf 79}, 3569 (1997).


\bibitem{smit} B. Smit and T.L.M. Maesen, Nature {\bf 451}, 06552 (2008).
%
\bibitem{smirnov} R.B. Gennis, in {\it Biophysical and  Structural Aspects of Bioenergetics}, M. Wikstr\"om ed. (RSC,  Cambridge, 2005).
%
\bibitem[{\citenamefont{Barker and Henderson}(1976)}]{barker76}
\bibinfo{author}{\bibfnamefont{J.~A.} \bibnamefont{Barker}} \bibnamefont{and}
  \bibinfo{author}{\bibfnamefont{D.}~\bibnamefont{Henderson}},
  \bibinfo{journal}{Rev. Mod. Phys.} \textbf{\bibinfo{volume}{48}},
  \bibinfo{pages}{587} (\bibinfo{year}{1976}).

\bibitem[{\citenamefont{Marconi and Tarazona}(1999)}]{marini99}
\bibinfo{author}{\bibfnamefont{U.M.B.} \bibnamefont{Marconi}}
  \bibnamefont{and} \bibinfo{author}{\bibfnamefont{P.}~\bibnamefont{Tarazona}},
  \bibinfo{journal}{J. Chem Phys.} \textbf{\bibinfo{volume}{110}},
  \bibinfo{pages}{8032} (\bibinfo{year}{1999});
  J.~Phys.: Condens.~Matter {\bf 12}, A413 (2000);
%
\bibinfo{author}{\bibfnamefont{A.J.} \bibnamefont{Archer}} \bibnamefont{and}
  \bibinfo{author}{\bibfnamefont{R.}~\bibnamefont{Evans}}, \bibinfo{journal}{J. Chem. Phys.} \textbf{\bibinfo{volume}{121}}, \bibinfo{pages}{4246}
  (\bibinfo{year}{2004}).

\bibitem[{\citenamefont{Evans}(1992)}]{evans1992fif}
\bibinfo{author}{\bibfnamefont{R.}~\bibnamefont{Evans}},
  \emph{\bibinfo{title}{Fundamentals of Inhomogeneous Fluids}}
  (\bibinfo{publisher}{New York, Dekker}, \bibinfo{year}{1992}).

\bibitem[{\citenamefont{Hansen and McDonald}(2006)}]{hansen2006tsl}
\bibinfo{author}{\bibfnamefont{J.~P.} \bibnamefont{Hansen}} \bibnamefont{and}
  \bibinfo{author}{\bibfnamefont{I.~R.} \bibnamefont{McDonald}},
  \emph{\bibinfo{title}{{Theory of Simple Liquids}}}
  (\bibinfo{publisher}{Academic Press}, \bibinfo{address}{London},
  \bibinfo{year}{2006}).

\bibitem[{\citenamefont{Percus}(1978)}]{percus76}
\bibinfo{author}{\bibfnamefont{J.~K.} \bibnamefont{Percus}},
  \bibinfo{journal}{J. Stat. Phys.} \textbf{\bibinfo{volume}{15}},
  \bibinfo{pages}{505} (\bibinfo{year}{1978}).

\bibitem[{\citenamefont{Doedel et~al.}(2001)\citenamefont{Doedel, Paffenroth,
  Champneys, Fairgrieve, Kuznetsov, Sandstede, and Wang}}]{AUTO}
\bibinfo{author}{\bibfnamefont{E.}~\bibnamefont{Doedel}~{\it et~al.}},
  \bibinfo{journal}{Technical Report, Caltech}  (\bibinfo{year}{2001}),
  \bibinfo{note}{url: http://cmvl.cs.concordia.ca/auto/};
%
\bibinfo{author}{\bibfnamefont{G.}~\bibnamefont{Bordyugov}} \bibnamefont{and}
  \bibinfo{author}{\bibfnamefont{H.}~\bibnamefont{Engel}},
  \bibinfo{journal}{Physica D} \textbf{\bibinfo{volume}{228}},
  \bibinfo{pages}{49} (\bibinfo{year}{2007}).

\bibitem{standard_ref}
See e.g.\ J.M.\ Brader and R.\ Evans, Physica A, {\bf 306} 287 (2002) and references therein.

\bibitem{taloni} A. Taloni and F. Marchesoni, Phys. Rev. Lett. {\bf 96}, 020601 (2006).

\bibitem[{\citenamefont{Savel'ev et~al.}(2003)\citenamefont{Savel'ev,
  Marchesoni, and Nori}}]{sav04}
\bibinfo{author}{\bibfnamefont{S.}~\bibnamefont{Savel'ev}},
  \bibinfo{author}{\bibfnamefont{F.}~\bibnamefont{Marchesoni}},
  \bibnamefont{and} \bibinfo{author}{\bibfnamefont{F.}~\bibnamefont{Nori}},
  \bibinfo{journal}{Phys. Rev. Lett.} \textbf{\bibinfo{volume}{91}},
  \bibinfo{pages}{010601} (\bibinfo{year}{2003});
  \bibinfo{journal}{Phys. Rev. E} \textbf{\bibinfo{volume}{71}},
  \bibinfo{pages}{011107} (\bibinfo{year}{2005}).


\end{thebibliography}
\end{document}